\documentclass[%
 aip,
 amsmath,amssymb,
 reprint,%
]{revtex4-1}

\usepackage{graphicx}% Include figure files
\usepackage{dcolumn}% Align table columns on decimal point
\usepackage{bm}% bold math
\usepackage[utf8]{inputenc}
\usepackage[T1]{fontenc}
\usepackage{mathptmx}
\usepackage{subfigure}
\usepackage{braket}
\usepackage{hyperref}
\usepackage{amsmath} % used for boldsymbol.
\hypersetup{colorlinks,citecolor=blue, linkcolor=blue, urlcolor=black}

\begin{document}

\preprint{AIP/123-QED}

\title{Thermal-demagnetization-enhanced hybrid fiber-based thermometer coupled with nitrogen-vacancy centers}
% Force line breaks with \\

\author{Shao-Chun Zhang}
\author{Shen Li}
\email{listen@mail.ustc.edu.cn}
\author{Bo Du}
\author{Yang Dong}
\author{Yu Zheng}
\author{Hao-Bin Lin}
\author{Bo-Wen Zhao}
\affiliation{CAS Key Laboratory of Quantum Information, University of Science and Technology of China, Hefei 230026, China}%
\affiliation{Synergetic Innovation Center of Quantum Information and Quantum Physics, University of Science and Technology of China, Hefei 230026, China}%
\author{Wei Zhu}
\author{Guan-Zhong Wang}
\affiliation{Hefei National Laboratory for Physical Science at Microscale, and Department of Physics,
University of Science and Technology of China, Hefei, Anhui 230026, P. R. China}
\author{Xiang-Dong Chen}
\author{Guang-Can Guo}

\author{Fang-Wen Sun}
\email{fwsun@ustc.edu.cn}
\affiliation{CAS Key Laboratory of Quantum Information, University of Science and Technology of China, Hefei 230026, China}%
\affiliation{Synergetic Innovation Center of Quantum Information and Quantum Physics, University of Science and Technology of China, Hefei 230026, China}%

\date{\today}% It is always \today, today,

\begin{abstract}
Nitrogen-vacancy centers in diamond are attractive as quantum sensors owing to their remarkable optical and spin properties under ambient conditions. Here we experimentally demonstrated a hybrid fiber-based thermometer coupled with nitrogen-vacancy center ensemble and a permanent magnet, where the temperature sensitivity was improved by converting the temperature variation to the magnetic field change based on the thermal-demagnetization of the permanent magnet. We have achieved both large temperature working range (room temperature to $373$ K) and millikelvin sensitivity ($1.6$ mK$/\sqrt{\rm{Hz}}$), nearly 6-fold improvement compared with conventional technique. This stable and compact hybrid thermometer will enable a wide range of applications for large-area detection and imaging with high temperature sensitivity.
\end{abstract}

\maketitle

%\begin{quotation}
%The ``lead paragraph'' is encapsulated with the \LaTeX\
%\verb+quotation+ environment and is formatted as a single paragraph before the first section heading.
%(The \verb+quotation+ environment reverts to its usual meaning after the first sectioning command.)
%Note that numbered references are allowed in the lead paragraph.
%%
%The lead paragraph will only be found in an article being prepared for the journal \textit{Chaos}.
%\end{quotation}

%\section{Introduction}

%Sensitive probing of temperature is an outstanding challenge in many areas of modern science and technology\cite{kucsko2013nanometre}.
A stable and compact thermometer capable of millikelvin resolution over a large temperature range could provide a powerful tool in many areas of physical, chemical, and biological researches\cite{kucsko2013nanometre}. Lots of promising approaches to local temperature sensing are being explored at present, including Raman spectroscopy\cite{kim2006micro}, scanning probe microscopy\cite{yue_nanoscale_2012}, and fluorescence-based measuremnets\cite{okabe2012intracellular} using nanoparticles and organicdyes\cite{vetrone2010temperature}. However, many of these methods are limited by drawbacks such as low sensitivity and systematic errors due to fluctuations in the fluorescence rate and the local environment\cite{yan2018coherent,kucsko2013nanometre}.

In recent years, the negatively charged nitrogen-vacancy (NV) center, a point defect in diamond, provides a promising system to realize practical quantum devices which have been successfully applied to a wide range of applications in quantum information processing and sensing in both physical and life sciences\cite{schirhagl2014nitrogen}. These applications of the NV center are based upon its remarkable optical and spin properties: bright optical fluorescence, long-lived spin coherence, and mature optical polarization and readout at room temperature\cite{yang2018solid}. For the NV-based temperature sensing, the techniques with modified spin-echo sequence\cite{kucsko2013nanometre} and high-order Carr-Purcell-Meiboom-Gill method\cite{wang2015high,toyli2013fluorescence} have achieved a sensitivity of $10$ mK$/\sqrt{\rm{Hz}}$. A nano-thermometer composed of NV centers and a magnetic nanoparticle has been experimentally demonstrated\cite{wang2018magnetic}, where an optimal temperature sensitivity of $3$ mK$/\sqrt{\rm{Hz}}$ has been obtained by the critical magnetization of the magnetic nanoparticle near Cuire temperature. Moreover, the recently developed fiber-optic probes coupled with NV centers were shown to enable a temperature measurement with a $20$ mK accuracy using optically detected magnetic resonance (ODMR)\cite{fedotov2014fiber,blakley2016fiber,safronov2015microwave}.

In order to further enhance the sensitivity of the NV thermometer for practical application, here, we proposed a hybrid fiber-based thermometer coupled with NV center ensembles and a permanent magnet. By converting the temperature variation to a magnetic field change\cite{wang2018magnetic,wojciechowski2018precision} of the permanent magnet, this thermometer can achieve a high sensitivity of $1.6$ mK$/\sqrt{\rm{Hz}}$ and a large temperature working range, where the permanent magnet is served as a transducer and amplifier of the local temperature variation owing to its temperature-dependent magnetisation\cite{wang2018magnetic,broadway2018high}.

\begin{figure}[b]
\centering
\includegraphics[width=0.47\textwidth]{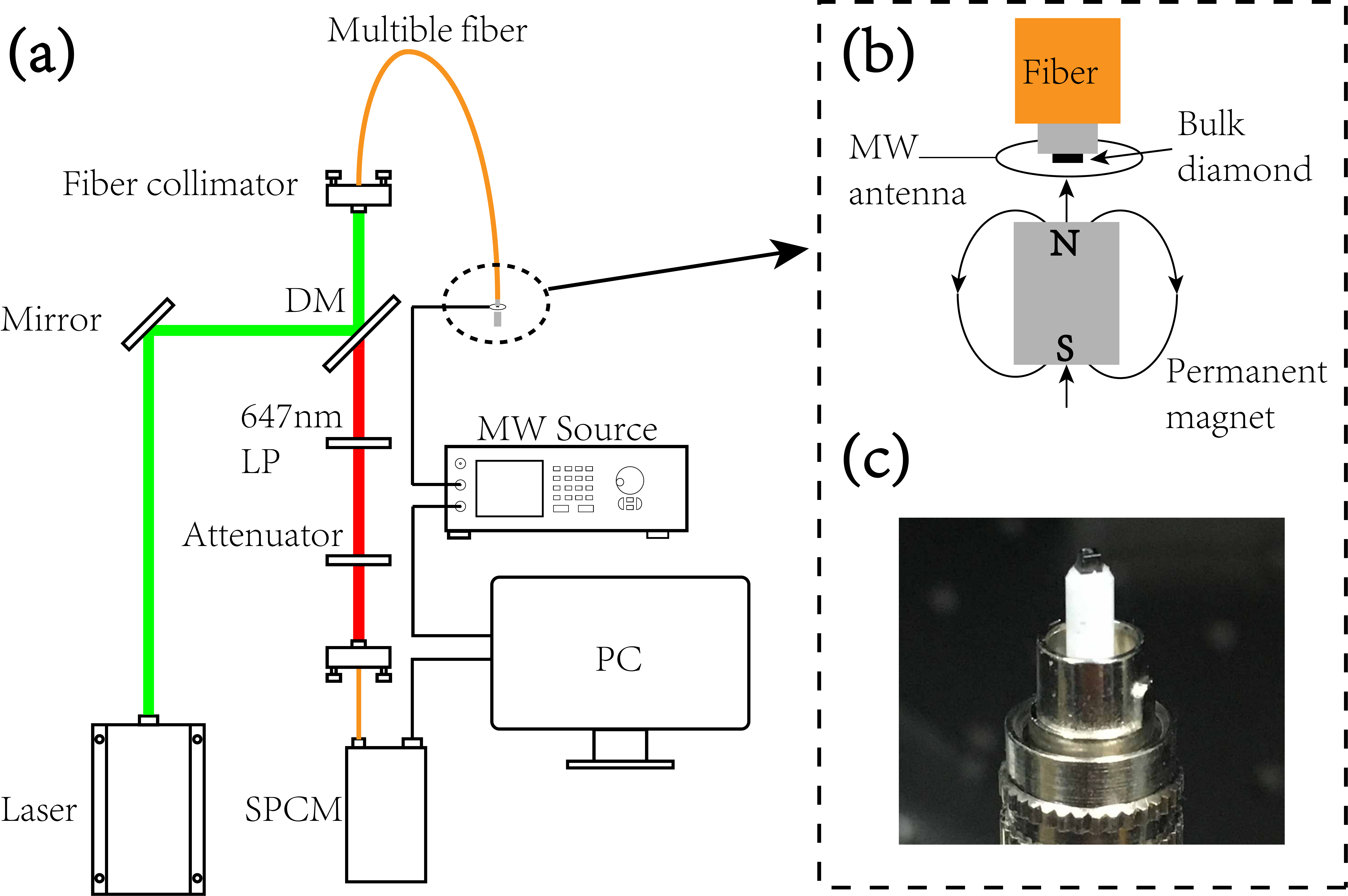}
\caption{\label{fig:experimentsetup}(a) The schematic of hybrid fiber-optical thermometer setup. SPCM, single photon counting module; DM, long pass dichroic mirrors with edge wavelength of $536.8$ nm. (b) A single crystal bulk diamond was attached on the tip of a multi-mode optical fiber. The cylindrical permanent magnet provided a magnetic field along [100] crystallographic direction. (c) Picture of the bulk diamond attached to the tip of a multi-mode optical fiber.}
\end{figure}

We used a homebuild fiber system and a microwave system to excite and detect the NV centers. A bulk diamond with the size of $1 \times$1$\times$1 mm$^3$ was attached on the tip of a multi-mode optical fiber with a core diameter of $100$ $\mu$m using UV curing glue, as shown in Fig.\ref{fig:experimentsetup}. The NV center ensembles in diamond (100) grown by plasma assisted chemical vapor deposition consisted of a [$N$] $\approx$ $40$ ppm and [$NV^-$] $\approx$ $0.15$ ppm. %Our experiment setup was similar to the confocal microscopy scheme except that objective lens is replaced by a multimode optical fiber [as shown in Fig.\ref{fig:experimentsetup}(a)].
In the experiment, the fiber delivered the $532$ nm laser to excite NV centers. Collected by the same fiber, the photoluminescence (PL) from the NV center ensemble passed through a $647$ nm long pass filter and an attenuator. Finally, it was detected by single photon counting module. Microwave was delivered by a printed circuit board with optical fiber fixed on it. The cylindrical Neodymium-Iron-Boron (NdFeB) permanent ($3$ mm $\times$ 15 mm) provided a bias magnetic field along the [100] axis of the diamond, as shown in Fig.\ref{fig:experimentsetup}(b) and (c). The magnetic field was projected equally onto all four orientations, resulting in a two-dip high contrast ODMR signal. The permanent magnet has a specified temperature coefficient $\alpha_0$ of the magnetisation $M(T)$ around room temperature which can be defined as
\begin{equation}\label{eq:magnetisation}
\alpha_0=\frac{1}{M(T)}\frac{\partial{M(T)}}{\partial{T}}\approx\frac{1}{B(T)}\frac{\partial{B(T)}}{\partial{T}},
\end{equation}
where $B(T)$ is the magnetic field, and $\alpha_0$ can be regarded as a constant at room temperature to 373 K for this type of permanent magnet\cite{wang2018magnetic,broadway2018high,calin2011temperature,yan2011preparation,sebastian1995temperature}.
%= -0.12 \pm{0.1} \%/$K\cite{wang2018magnetic,broadway2018high}.

The negatively charged NV center in diamond consists of a substitutional nitrogen associated with a vacancy in an adjacent lattice site of the diamond crystal. This defect exhibits an efficient and photostable red PL, which enables optical detection at room temperature\cite{suter2017single}. The ground state is a spin triplet with $^3A_2$ symmetry including a singlet state $m_s = 0$ and a doublet state $m_s = \pm{1}$ separated by a temperature-dependent zero-field splitting (ZFS) $D=2.87$ GHz in the absence of magnetic field. Applying a static magnetic field along the NV axis leads to a splitting of $m_s = -1$ and $m_s = +1$ states. Moreover, the ground states are coupled to a spin triplet excited state $^3E_2$ using green light ($532$ nm)\cite{jensen2013light,dreau2011avoiding}.

Considering the temperature effect, the spin Hamiltonian of the ground state\cite{tetienne2012magnetic} can be written as
\begin{equation}\label{eq:Hamiltonian}
H_{NV}\approx D(T)S_{z}^{2}+E(S_x^2-S_y^2)+\gamma_e\textbf{B}(T)\cdot\textbf{S} \text{.}
\end{equation}
Here, \textbf{S} is the electronic spin operator. $E$ represents the local strain in the diamond which is almost temperature-independent\cite{acosta2010temperature}. \textbf{B}$(T)$ is the applied magnetic field, % determined by magnetization \textbf{M(T)} of the permanent magnet,
and $\gamma_e = 2.8$ MHz/G. Assuming that $D \gg \gamma_e B$, the changing temperature $\delta T$ of the diamond results in a transition frequencies shift $\delta f$, which can be described as
\begin{equation}\label{eq:sensitivity}
\delta f\approx \{\frac{\partial{D(T)}}{\partial{T}}+[3(\gamma_e\rm{sin}(\theta))^2\frac{\emph{B(T)}}{\emph{D(T)}}\pm\gamma_e\rm{cos}(\theta)]\cdot\frac{\partial{\emph{B(T)}}}{\partial{\emph{T}}}\}\delta\emph{T}\text{,}
\end{equation}
where $\theta$ is the angle between magnetic field $B$ and NV axis. ${\partial{\emph{B(T)}}}/{\partial{\emph{T}}}$ is the responsibility to the temperature shifts of the permanent magnet.
%,the '-' and '+' corresponds to the electron transition $\ket{m_s=0}\leftrightarrow\ket{m_s=-1}$ and $\ket{m_s=0}\leftrightarrow\ket{m_s=+1}$ .

\begin{figure}[t]
\centering
%\flushleft
\includegraphics[width=0.4\textwidth]{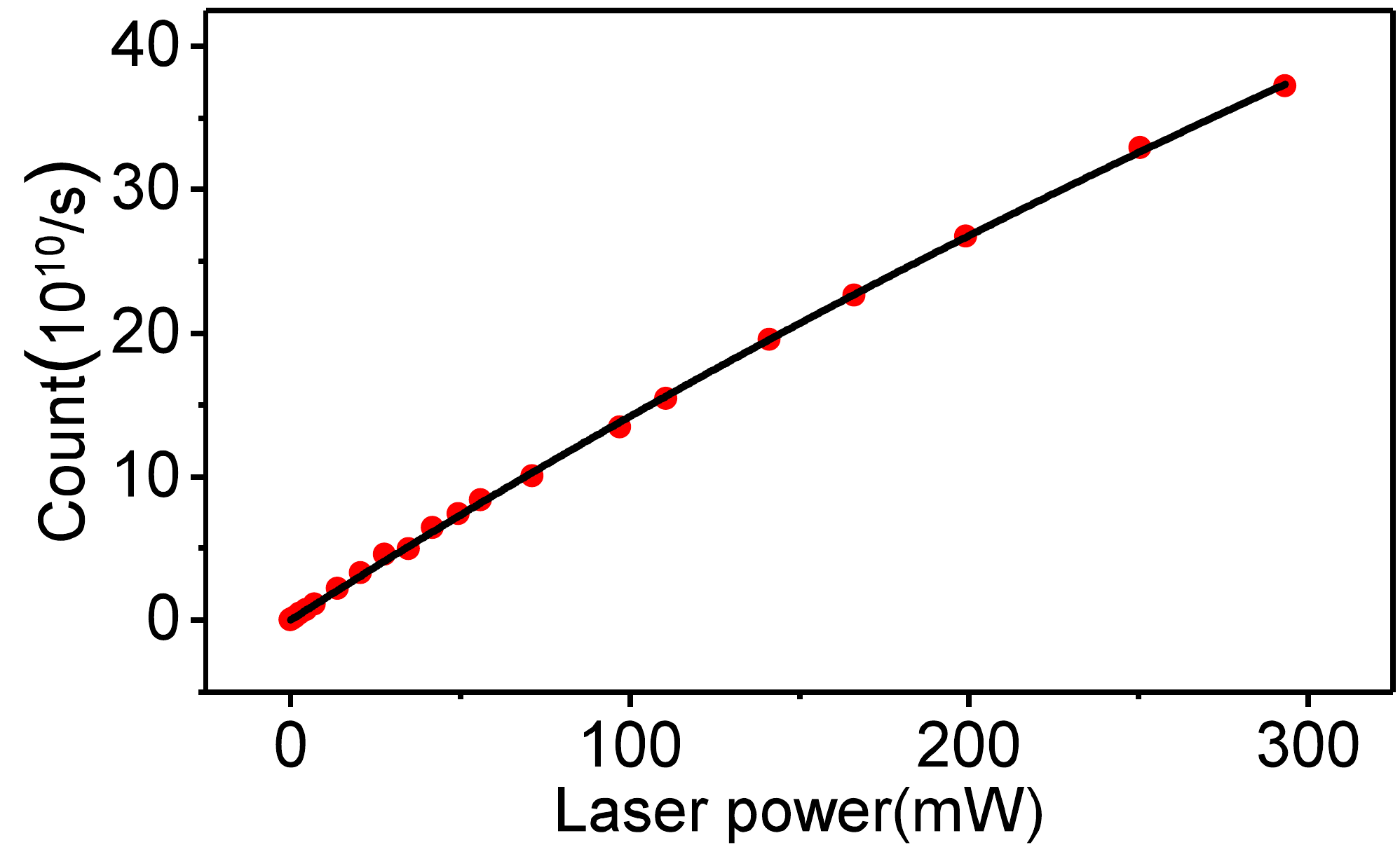}
\caption{\label{fig:fluorescence}Measured fluorescence $P_{\rm{fl}}$ as a function of pump power $P$. Solid line is a fit to the function of $P_{\rm{fl}} = kP/(1+P/P_{\rm{sat}})$.}
\end{figure}

In the experiment, we executed the temperature measurements with two schemes. The first measurement was performed in the absence of the bias magnetic field ($B$ = 0 G). It focused on the precise measurement of the ZFS shift with temperature. Hence Eq.\eqref{eq:sensitivity} becomes $\delta f\approx \frac{\partial{D(T)}}{\partial{T}}\delta T$. In the second measurement, we applied a temperature-dependent bias magnetic field along the [100] axis of the diamond where the angle between the magnetic field and each NV axis is $\theta = 54.7^\circ$ \cite{tetienne2012magnetic,fukui2014perfect}.

According to Eq.\eqref{eq:Hamiltonian} and Eq.\eqref{eq:sensitivity}, the temperature sensitivity $\delta T$ is limited by the resonant frequency resolution $\delta f$ of the ground sub-levels transition, which is similar to the sensitivity of DC magnetic field measurement. The principle of the magnetic field measurement has been well demonstrated\cite{degen2017quantum}. The sensitivity has been analyzed theoretically and experimentally from Ramsey pulse sequences or pulsed-ODMR measurement\cite{degen2017quantum,rondin2014magnetometry,chen2013vector,dong2018non}. Although ultrahigh sensitivity can be achieved by these techniques, the simplest way to detect an external DC magnetic field with NV ensemble remains the direct evaluation of ODMR\cite{hayashi2018optimization}, especially in the fiber-based diamond sensing for practical application\cite{liu2013fiber,dong2018fiber}. The shot-noise-limited sensitivity of the ODMR measurement is linked to the resonant frequency resolution $\delta f$, which is read as
\begin{equation}\label{eq:resolution}
\delta f\approx P_{F}\frac{\Delta\nu}{C\sqrt{I_0}} \text{,}
\end{equation}
where $I_0$ is the rate of detected photons per unit of time, $C$ is the ODMR contrast associated to the dip of PL intensity, $P_F$ is the parameter of the line shape and $\Delta\nu$ is the linewidth\cite{degen2017quantum,rondin2014magnetometry}.

%\section{Results}

\begin{figure}[t]
\centering
\includegraphics[width=0.47\textwidth]{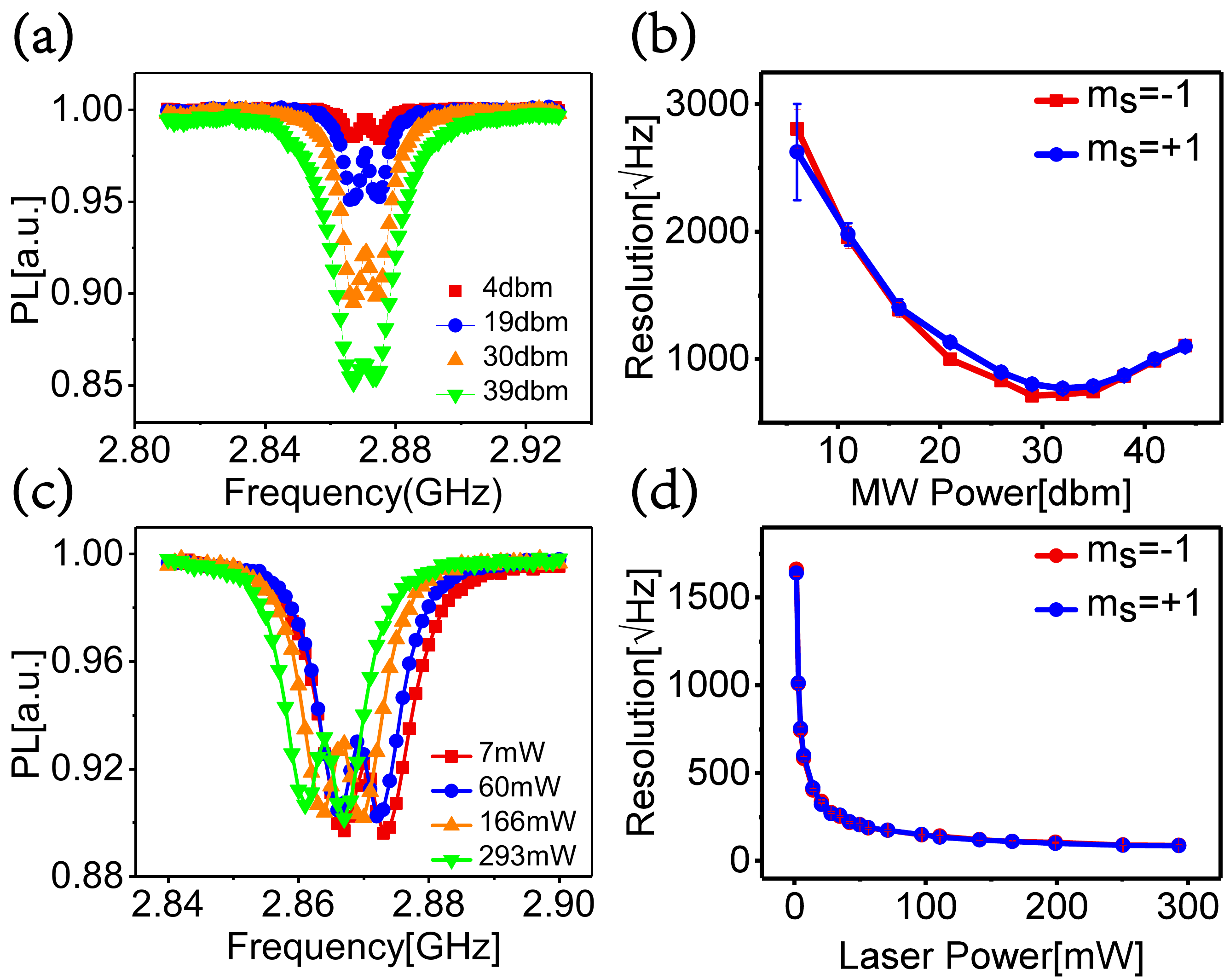}
\caption{\label{fig:MW}Examples of ODMR signals and the resonant frequency resolution. (a) ODMR signals with different settings of the microwave powers. The pump laser power was fixed to $7$ mW. (b) The resonant frequency resolution as a function of the microwave power. The red and blue lines represent the electron transition $\ket{m_s = 0} \leftrightarrow \ket{m_s = -1}$ and $\ket{m_s = 0} \leftrightarrow \ket{m_s = +1}$, respectively. (c) ODMR signals with different settings of pump laser power. Microwave power was fixed to $30$ dbm. (d) The resonant frequency resolution as a function of pump laser power.}
\end{figure}

In the experiment, we first studied the conditions of the best resonance frequency resolution of the ODMR signals, including the laser power and microwave power\cite{jensen2013light,dreau2011avoiding}. The amount of red fluorescence as a function of green light power was measured, which is plotted in Fig.\ref{fig:fluorescence}, together with a fit of the form $P_{\rm{fl}} = kP/(1+P/P_{\rm{sat}})$\cite{jensen2013light}. Then, by fixing the laser light power $P$ to $7$ mW, we detected the ODMR signals with different settings of the microwave power ${P}_{\rm{MW}}$, as shown in Fig.\ref{fig:MW}(a). Even in the absence of external magnetic field, the local strain removes the degeneracy of the ground sublevels $m_s = \pm{1}$, giving rise to two well-resolved features in ODMR spectra. We can clearly observed that the increasing of the MW power leads to the broadening of the ODMR, as well as the increasing contrast. From this set of measurements, the resonant frequency resolution $\delta f$ as a function of microwave power can be estimated using Eq.\eqref{eq:resolution}. An optimal resolution $\delta f\approx710$ $\sqrt{\rm{Hz}}$ can be obtained with a typical microwave power $P_{\rm{MW}}\approx 30$ dbm, %corresponding to one of the conditions to get the best resolution,
as shown on Fig.\ref{fig:MW}(b).

%Then we proceed the study of the condition of pump laser power.
By keeping $P_{\rm{MW}}$ fixed to $30$ dbm, we experimentally measured the ODMR spectrum for pump laser power $P$ ranging from $0$ to $293$ mW, as shown in Fig.\ref{fig:MW}(c). The linewidths extracted from the spectra show a decrease with the increase of pump power, dues to the effect of light-narrowing\cite{jensen2013light}. The pump laser with increasing power heats the diamond at the end of optical fiber, resulting in a shift of the spectra which agrees with the measurement in earlier study\cite{fedotov2014fiber}. Moreover, the laser-heating effect further leads to a decreasing contrast\cite{liu2018quantum}. According to Eq.\eqref{eq:resolution} for the resonant frequency resolution, the best resolution $\delta f\approx87$ $\sqrt{\rm{Hz}}$ was reached for microwave power $P_{\rm{MW}} = 30$ dbm and light power $P = 293$ mW, as shown in Fig.\ref{fig:MW}(d). However, the resolution is expected to become worse eventually for higher light power.

\begin{figure}[tbp]
\centering
\includegraphics[width=0.47\textwidth]{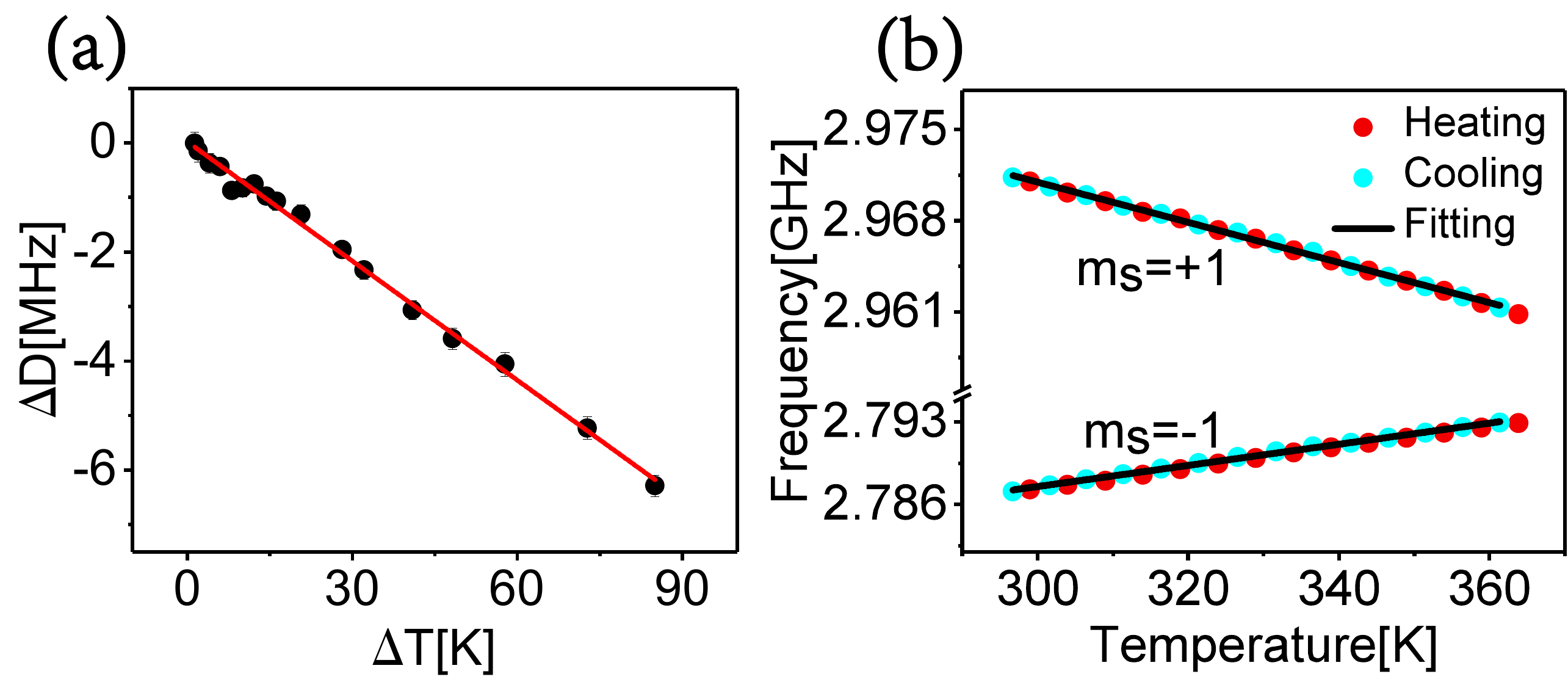}
\caption{\label{fig:temperature}The resonant frequency shifts resulting from the temperature. (a) The measured ZFS parameter $D$ as a function of temperature ranging from room temperature to 373K in the absence of the bias magnetic field. The red line is the theoretical fit with a function from Refs.[\onlinecite{chen2011temperature},\onlinecite{li2017temperature}]. (b) The resonant frequency of the ODMR as a function of the temperature of permanent magnet by keeping the diamond with constant temperature. Both heating (red dots) and cooling (cyan dots) processes were measured. With linear fit, the two slops were estimated to be $k_1 = -155(2)$ $\rm{kHz}/\rm{K}$ and $k_2 = 90(1) \rm{kHz}/\rm{K}$, respectively.}
\end{figure}

For NV-based temperature measurement, we first recorded the ODMR spectra with the diamond at temperatures range from $293$ K to $373$ K. To control the diamond's temperature, the diamond was mounted to a flexible resistive foil heater (HT10K, Thorlabs). The heater and the resistive temperature detector were both controlled by a temperature controller (TC200, Thorlabs) to achieve the temperature stability within $\pm{0.1}$ K up to $373$ K in an atmospheric environment. However, the laser heating can significantly affect the detection accuracy of temperature. Firstly, we set the laser light power to $7$ mW rather than $293$ mW. In this case, the temperature  of the diamond can be kept at room temperature, indicating that the heating effect can be ignored and the local temperature variation can be transferred to the quantum sensor. In the measurement process, the ZFS parameter $D$ was detected as a function of temperature, which is plotted and fitted in Fig.\ref{fig:temperature}(a). It shows a linear decrease with a slope of $-74(1)$ $\rm{kHz}/\rm{K}$. The optimal temperature sensitivity $\delta T$ using Eq.\eqref{eq:sensitivity} was estimated to be $9.4$ $\rm{mK}/\sqrt{\rm{Hz}}$. However, the ZFS parameter $D$ for NV center presents only a weak dependence on temperature (a small ${\partial{D(T)}}/{\partial{T}}$). Moreover, the laser pump power can not be set at a higher level for more photon counts. Both of the these restrictions limit the further improvement of the sensitivity.

On the other hand, NV centers have been confirmed to be ultra-sensitive to external magnetic field \cite{barry2019sensitivity}. Here, we demonstrated another scheme to improve the sensitivity by converting the temperature variation to a magnetic field change\cite{wang2018magnetic,liu2018quantum}. We applied a temperature-dependent magnetic field along the [100] axis of diamond by a permanent magnet. %Here, we need first to test the temperature coefficient $\alpha_0$ of permanent magnet.
Under the conditions of the best resonance frequency resolution with the highest photon rate from Fig\ref{fig:MW}(d), the bulk diamond on the tip of fiber was irradiated by laser with a constant power of $293$ mW which kept the diamond with a constant temperature\cite{fedotov2014fiber}, $T_0\approx 400$ K. With a heat insulation treatment, heating the permanent magnet will not induce the temperature shift of diamond. In this case, the temperature variation was converted to the magnetic field change which was detected by the NV center through the ODMR measurement. In the experiment, the ODMR measurement were carried out for temperatures of the permanent magnet from $293$ K to $373$ K, and the correlation between NV sublevels resonant transition frequency $\omega_{\pm}$ and temperature are shown in Fig.\ref{fig:temperature}(b). By linear fitting, the two slops were estimated to be $k_1 = -155(2)$ $\rm{kHz}/\rm{K}$ and $k_2 = 90(1) \rm{kHz}/\rm{K}$, respectively. Extracting the applied magnetic field $B$ from Fig.\ref{fig:temperature}(b), the temperature coefficient $\alpha_0$ of permanent magnet was estimated to be $-0.14(1)\%/K$ by Eq.\eqref{eq:magnetisation}. Moreover, the frequency shift induced by the magnetization of the permanent magnet is reversible when the temperature is scanned back, as shown in Fig.\ref{fig:temperature}(b) with cyan dots, which indicates the stability of this hybrid sensor\cite{wang2018magnetic}.

\begin{figure}[tbp]
\centering
\includegraphics[width=0.48\textwidth]{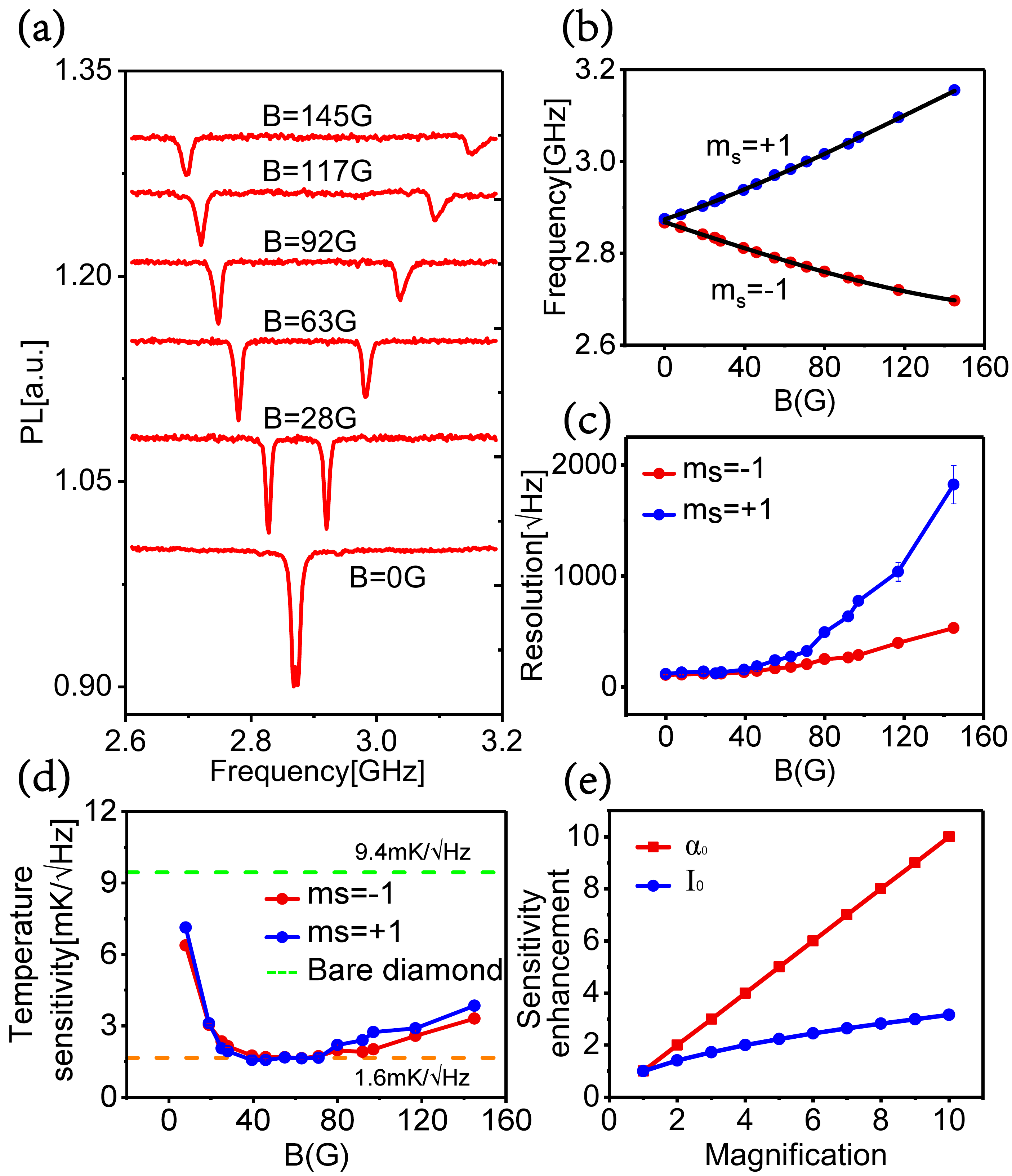}
\caption{\label{fig:max sensitivity}(a) Examples of ODMR with different settings of bias magnetic field $B$. (b) The resonant frequency extracted from (a) as a function of magnetic field $B$. Solid line is the fit using Eq.\eqref{eq:Hamiltonian}. (c) The estimated resonant frequency resolution $\delta$f as a function of $B$. (d) Plot of the temperature sensitivity with bias magnetic field. The optimal sensitivity of this hybrid thermometer demonstrates an almost $6$-fold improvement compared with conventional technique with bare diamond (dashed green line). (e) The temperature sensitivity enhancement via the  improvement of photon counts $I_0$ and temperature coefficient $\alpha_0$ of the permanent magnet.}
\end{figure}

According to Eq.\eqref{eq:magnetisation} and Eq.\eqref{eq:sensitivity}, both ${\partial{B(T)}}/{\partial{T}}$ and the resonant frequency resolution $\delta f$ contribute to the temperature sensitivity $\delta T$. To achieve the best temperature sensitivity of this hybrid sensor, ${\partial{B(T)}}/{\partial{T}}\approx\alpha_0B$ should be large and $\delta f$ should be the smallest. Fig.\ref{fig:max sensitivity}(a) shows ODMR spectra with different settings of magnetic field $B$ and the extracted electron spin transition frequencies are plotted in Fig.\ref{fig:max sensitivity}(b). However, the contrast of the ODMR signals significantly decreased with the increasing $B$ since the magnetic field was not along the NV axis\cite{tetienne2012magnetic} exactly, which led to a low resonant frequency resolution eventually, as shown in Fig.\ref{fig:max sensitivity}(c). In this case, there is an extremum point for a typical magnetic field to obtain optimal temperature sensitivity. Given by Eq.\eqref{eq:sensitivity}, the sensitivity $\delta T$ as a function of magnetic field $B$ is plotted in Fig.\ref{fig:max sensitivity}(d). The optimal sensitivity of $1.6$ mK$/\sqrt{\rm{Hz}}$ was reached for the magnetic field at the range of $40-60$ G. In comparison with the result of the bare bulk diamond on the tip of fiber, as shown in Fig.\ref{fig:max sensitivity}(d), the sensitivity can be almost improved by $6$-fold of magnitude using the hybrid thermometer. This scheme will become ultra-sensitive when the working temperature close to the magnetic phase transition point of the permanent magnet for a large temperature coefficient $\alpha_0$\cite{wang2018magnetic}. However, it will narrow the working range.

In the experiment, the sensitivity was limited by the density of NV centers. The generating efficiency of NV center ensemble is less than $1\%$ for our diamond sample. And by electron irradiation treatment, the density of NV center can be enhanced more than $30$ times\cite{xu2018room}. With high-density NV ensemble, the temperature sensitivity can be improved by a factor of $\sqrt{N}$\cite{rondin2014magnetometry} when the collected PL signal is magnified by the number $N$ of the sensing spins, as shown in Fig.\ref{fig:max sensitivity}(e) with blue line. A micro-concave mirror on the tip of the fiber can further improve the fluorescence excitation and collection\cite{duan2019efficient,duan2018enhancing}. Moreover, a ferromagnetic material with higher temperature coefficient of the magnetisation at room temperature, such as vanadium oxide\cite{krusin2004room} ($\alpha_0\approx -0.8\%$/K) and Ni-Mn-Sn alloys\cite{krenke2005inverse} ($\alpha_0\approx -1\%$/K), can also significantly enhance the sensitivity, as shown in Fig.\ref{fig:max sensitivity}(e) with red line. All of these methods can boost temperature sensitivity of this hybrid thermometer toward sub-$0.1$ mK$/\sqrt{\rm{Hz}}$ over a large temperature range.

In summary, we have demonstrated a fiber-based hybrid thermometer with NV center ensembles in a bulk diamond. Based on thermal-demagnetization effect, the permanent magnet was served as a transducer and amplifier of the local temperature variation. We have achieved the temperature sensitivity of $1.6$ mK$/\sqrt{\rm{Hz}}$ ranging from $293$ K to $373$ K. With further improvement on the sensitivity, such a stable and compact thermometer will be widely applied in physical, chemical, and biological science and technology.
%which provides a temperature-dependent magnetic field along the [100] axis, hence all the NV centers contribute the ODMR signals. We first obtain the conditions of the best frequency resolution of ODMR signals, including the laser light power and microwave power. Then we perform ODMR measurements with the temperature of diamond ranging from 293 K to 373 K, which shows a sensitivity of 9.4 mK$/\sqrt{\rm{Hz}}$ via the shift of the zero-field splitting D. Finally, we experimentally show that the hybrid scheme obtains more than a factor of 6 improvement sensitivity () than the scheme with a bare bulk diamond.

\section*{Acknowledgment}
This work is supported by the National Key Research and Development Program of China (No. 2017YFA0304504), the Science Challenge Project (No. TZ2018003), the National Natural Science
Foundation of China (Nos. 91536219, 61522508, and 91850102), the Anhui Initiative in Quantum Information Technologies (No. AHY130000).
%\cite{wang2018magnetic}
%\nocite{*}
\bibliographystyle{apsrev4-1}
%\bibliography{ref}% Produces the bibliography via BibTeX.
%

\end{document}